\documentstyle[aps,bezier,epsfig,amsmath,prb,graphicx,amssymb]{revtex}

\renewcommand{\epsilon}{\varepsilon}
\renewcommand{\phi}{\varphi}

\newcommand{\be}{\begin{equation}}
\newcommand{\ee}{\end{equation}}

\begin{document}

\draft
\twocolumn[\hsize\textwidth\columnwidth\hsize\csname @twocolumnfalse\endcsname








\vskip2pc]

\narrowtext

{\bf Comment on `Symmetrical Temperature-Chaos effect with Positive and
Negative Temperature Shifts in a Spin Glass'.}

In a very interesting paper, J\"onsson, Yoshino and 
Nordblad~\cite{Suedois-Yoshino1} have shown 
that the effect of small temperature shifts on the aging behaviour of 
an Heisenberg-like spin glass could be interpreted quantitatively 
using the ideas of temperature chaos and overlap length.
In this comment, we show that the same analysis can be performed 
in a case where temperature chaos is known to be irrelevant, 
weakening the main conclusion of Ref.~\cite{Suedois-Yoshino1}.

As now well established, aging at low temperatures $T$
in spin glasses is associated with the slow 
growth with time $t$ of a coherence length, $\ell_T(t)$. This length can be 
measured in simulations, but can also be inferred from experimental 
data, using plausible assumptions, leading to a rather consistent 
determination of $\ell_T(t)$. ``Cumulative aging''~\cite{Suedois-Yoshino1}
means that the same coherence length grows at different
temperatures, although at different rates. In this case, the value of 
$\ell_{T_i}(t_w)$ after staying a certain time $t_w$ 
at a first temperature $T_i$ serves as the
`initial condition' for the growth of $\ell$ after a temperature 
shift. 

Rejuvenation effects, on the other hand, demonstrate that 
cumulative aging cannot be the only story in spin glasses.
The ``temperature chaos'' scenario 
postulates that typical equilibrium configurations in a spin glass at 
two temperatures differing by $\Delta T$ are strongly correlated only up to
the ``overlap length'', $\ell_{\Delta T}$, beyond which 
these correlations rapidly
decay to zero. From scaling arguments, one expects 
$\ell_{\Delta T} \sim |\Delta T|^{-1/\zeta}$, 
with $\zeta \approx 1$ for the Ising spin glass. If this 
scenario holds, one expects that the initial condition  
for the growth of $\ell$ after a temperature shift, as
encoded by an effective
length $\ell_{\rm eff}$,
will obey the following scaling form:
\begin{equation}
\label{scaling}
\frac{\ell_{\rm eff}}{\ell_{\Delta T}} = F\left(
\frac{\ell_{T_i}(t_w)}{\ell_{\Delta T}}
\right),
\end{equation}
with $F(x \ll 1)=x$ (cumulative aging) and $F(x \gg 1)=1$ (temperature chaos).
Rejuvenation, i.e. 
deviations from cumulative aging, can thus be accounted for by 
temperature chaos.
The verification of Eq.~(\ref{scaling}) using experimentally 
determined values of $\ell_{T_i}$ and $\ell_{\rm eff}$ for different 
$\Delta T$ and $t_w$ is the central 
result of Ref.~\cite{Suedois-Yoshino1}, thereby
providing strong support for the temperature chaos scenario and a numerical
value for the chaos exponent for the AgMn spin-glass, $1/\zeta \approx 2.6$. 
In the above argument, one assumes from the start that rejuvenation is 
induced by temperature chaos, and finds {\it self-consistent} results. 

However, 
other scenarii have been proposed in the literature to explain rejuvenation 
effects, such as the progressive freezing of smaller and smaller 
length scale modes~\cite{BB,JP}. In this respect, we recently 
demonstrated in a numerical simulation of the 4d Ising spin glass that strong 
rejuvenation effects can be observed 
in conditions where the overlap length is {\it independently}
observed to be much larger than all relevant length scales~\cite{BB}.

\begin{figure*}
\begin{center}
\psfig{file=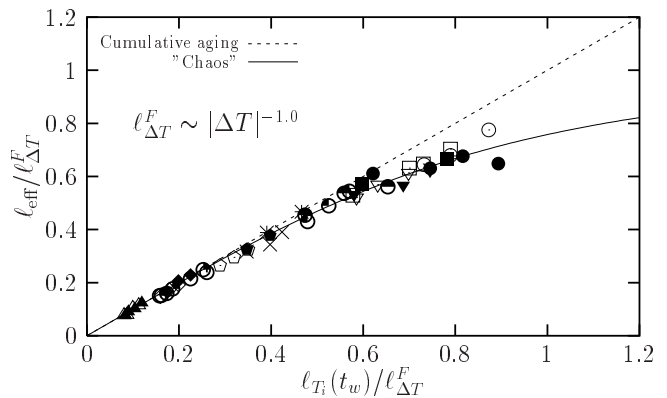,height=5.3cm}
\end{center}
\caption{Test of Eq.~(\ref{scaling}) for a series of ``twin experiments''
in the 4d Gaussian Ising spin glass
strictly following protocols and analysis of Ref.~\cite{Suedois-Yoshino1}.
56 shift experiments are presented with $T_i/T_c \in [0.4,0.9]$,
$\Delta T/T_c = 0.05$, 0.1, 0.2, $\cdots$, 0.5, $t_w \in [80,57797]$.
System size is $L=25 \gg \ell_T(t)$.
$\ell_{\rm eff}$ was defined from the maximum of $\partial 
C(t+t_w,t_w)/\partial t$, where $C$ is 
the spin autocorrelation function (instead
of TRM) and growth
laws $\ell_T(t)$ taken from Ref.~\cite{BB}.
The exponent $\zeta$ of the fictitious overlap length $\ell_{\Delta T}^F$
is chosen to collapse {\it all} our data.}
\label{4point}
\end{figure*}

To understand these contradictory results~\cite{Suedois-Yoshino1,BB},  
we have reproduced the protocol of 
Ref.~\cite{Suedois-Yoshino1} in an extensive new series of simulations, 
and followed the very same steps to determine the length scales 
$\ell_{T_i}(t_w)$ and $\ell_{\rm eff}$. 
Although we do know that all dynamic length scales in our simulations 
are $\leq 5$ and that the overlap length is much larger
(probably larger than $20$), 
we tried to rescale all our results using Eq.~(\ref{scaling}). 
Surprisingly, as shown in Fig.~\ref{4point}, 
this works very well with the expected value $\zeta \approx 1$. 
Rejuvenation results then in our case in the 
appearance of a {\it fictitious} overlap length, $\ell_{\Delta T}^F$.
This shows that although suggestive, the
analysis of Ref.~\cite{Suedois-Yoshino1} cannot be viewed as definitive 
evidence for temperature chaos. 

Simulations were performed on OSWELL 
at the Oxford Supercomputing Center, Oxford University, UK.

\vspace*{.3cm}

Ludovic Berthier$^1$ and Jean-Philippe Bouchaud$^2$

$^1${Theo. Phys., 1 Keble Road, Oxford, OX1 3NP, UK.}

$^2${SPEC,
Orme des Merisiers, 91191 Gif/Yvette, France.}

\end{document}